# Rigorous Theory of the Thin Vapor Layers Optical Properties: The Case of Specular Reflection of Atoms Colliding with the Walls

A. V. Ermolaev and T. A. Vartanyan

*ITMO University, Kronversky pr. 49, 197101 St. Petersburg, Russia*

The theory of the thin vapor layers linear optical properties is presented for the case of specular reflection of atoms colliding with the walls. The effects of light absorption and the shift in the resonance frequency are taken into account by means of self-consistent calculation of the field and polarization in a gaseous medium. The obtained formulas reproduce the complex dependence of the spectral line profile on the gas layer thickness and allow one to determine numerically the exact values of the "blueshift".

## I. INTRODUCTION

The possibility of observing sub-Doppler resonances in the spectra of light reflection from the interface between a transparent dielectric medium and atomic vapor makes selective reflection spectroscopy a promising area for theoretical and experimental studies. Doppler-free structures in the line shape of selective reflection were first observed by Cojan [1]. In this work, the narrowing of the spectral lines in reflection was ascribed to the transient polarization of the atoms departing from the cell wall. In the limit of sufficiently high vapor densities, particle collisions lead to a rapid loss of the oscillation phase. As a consequence, polarization retains local connection with the field. Contrary to that, in the case of rarefied vapors when the mean free path of atoms without loss of polarization becomes greater than the wavelength of the incident light spatial dispersion effects cannot be neglected [2].

Recently, it was demonstrated [2,3] that the nonlocal optical response of a resonant gas can be significantly enhanced if the vapor is spatially confined in a cell with a thickness of the order of the incident wavelength $\lambda$. The invention of a gas cell with a subwavelength thickness [4] enabled experimenters to examine optical properties of a thin vapor layers, as well as the interaction of atoms with surfaces of dielectric materials [5,6]. These studies may become the basis for the creation of miniature atomic standards of frequency and time [7]. From this point of view, it is important to correctly consider factors leading to a shift and broadening of the spectral lines of selective reflection. In this article we will focus on one of the shifts of a purely electrodynamics nature that arises due to the interference between the contributions of the departing and arriving atoms (hereinafter referred to as the "blueshift") [8].

In general, an accurate theoretical description of the interaction of light with a thin layer of resonant gas presents significant mathematical difficulties. In Ref. [3] the approximate solution for the reflected and transmitted fields was obtained by means of perturbation expansion of the field with respect to the atomic vapor density for the case of diffuse boundary conditions. This implies that the atom loses polarization upon collision with the wall. In fact, in order to account for the structure of the field inside the vapor exactly, the field and polarization inside the vapor should be calculated in a self-consistent way [9]. The goal of this study is to investigate the linear optical properties of thin vapor layers beyond the scope of perturbation theory. We will assume throughout this paper that atoms collide with the walls specularly, i. e. the atoms preserve their polarization and only the normal component of their velocity changes sign after a collision with the wall. In practice to some extent, this situation can be realized by applying anti-relaxation coatings on the windows of the gas cell [10].

## II. THEORETICAL MODEL

Consider a vapor layer of thickness $l$, located between two transparent dielectric media with parallel interfaces. The described system is schematically shown in Fig. 1, where a monochromatic electromagnetic plane wave propagates in the positive direction of the *x*-axis in the first dielectric medium with refractive index $n_1$. The wave (at normal incidence) partially reflects from the gas layer and penetrates into the second dielectric medium with refractive index $n_2$.

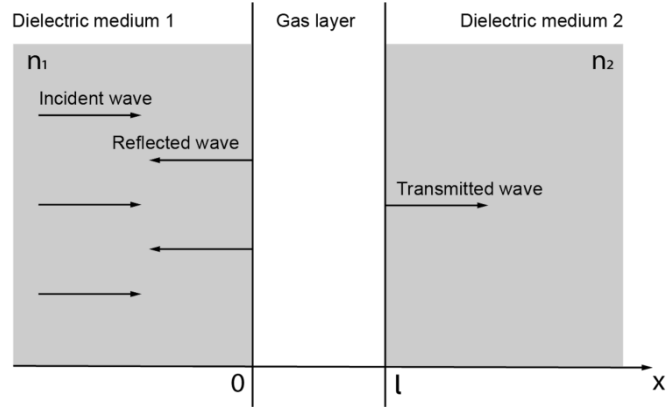

FIG. 1. Schematic illustration of light reflection (at normal incidence) from a thin gas layer

Neglecting nonlinear optical processes, the electric field and polarization inside the vapor can be represented as

$$E(x,t) = \frac{1}{2} E(x)\exp(-i\omega t) + c.c., \qquad (1a)$$

$$P(x,t) = \frac{1}{2} P(x)\exp(-i\omega t) + c.c.. \qquad (1b)$$

From the Maxwell's equations it can be shown that the spatial parts of the electric field and polarization of the medium satisfy the equation:

$$\frac{d^2 E(x)}{dx^2} + k^2 E(x) = -4\pi k^2 P(x), \qquad (2)$$

where $k = \omega/c$.

It is assumed that the gas consists of two-level atoms and the frequency of the incident light wave $\omega$ is varied in the vicinity of the atomic transition frequency $\omega_0$. Then, the macroscopic polarization can be expressed in terms of the off-diagonal density-matrix element $\rho(x,\upsilon)$ by averaging the dipole moment operator over the velocity distribution of atoms $f(\upsilon)$

$$P(x) = 2ND \int_{-\infty}^{+\infty} d\upsilon f(\upsilon) \rho(x,\upsilon), \qquad (3)$$

where $N$ is the number density of the gas atoms, $D$ is the transition dipole moment, and $\upsilon$ is the projection of the atomic velocity on the $x$-axis. Throughout this article we will assume that the thermal motion of the gas layer atoms is described by the Maxwell's distribution function

$$f(\upsilon) = \frac{1}{\sqrt{\pi}\upsilon_T} \exp\left(-\frac{\upsilon^2}{\upsilon_T^2}\right), \qquad (4)$$

where $\upsilon_T$ is the most probable thermal velocity.

In order to study the linear optical properties of the described system, we restrict ourselves to the case of small electric field strengths, when the intensity of the light is so low, that it cannot saturate the resonant transition. In the resonance region, the off-diagonal element of the density matrix $\rho(x,\upsilon)$ satisfies the equation:

$$\upsilon \frac{\partial \rho(x,\upsilon)}{\partial x} + \left[\gamma + i(\omega_0 - \omega)\right]\rho(x,\upsilon) = \frac{i}{2\hbar} DE(x), \tag{5}$$

where $\gamma$ is the homogenous width of the transition that is the sum of natural and collisional widths of the spectral line. After substituting the expression for polarization [see Eq. (3)] in Eq. (2), we can rewrite the whole set of Eqs. (2)-(5) in the following way:

$$\frac{d^2 E(\xi)}{d\xi^2} + E(\xi) = -2im \int_{-\infty}^{+\infty} \sigma(\xi, v) e^{-v^2} dv, \tag{6a}$$

$$v \frac{\partial \sigma(\xi, v)}{\partial \xi} + \eta \sigma(\xi, v) = E(\xi). \tag{6b}$$

Here we have introduced the dimensionless variables $\xi = kx$, $m = 2\sqrt{\pi} ND^2 / \hbar k \upsilon_T$, $v = \upsilon / \upsilon_T$, $\sigma(\xi, v) = (2\hbar k \upsilon_T / iD)\rho(x,\upsilon)$, $\eta = \Gamma - i\Omega$, $\Gamma = \gamma / k\upsilon_T$, and $\Omega = (\omega - \omega_0)/k\upsilon_T$, where parameter $m$ is proportional to the atomic vapor density.

For further consideration, we also need to impose boundary conditions for the field at the boundaries of the gas layer. In accordance with the statement of the problem [see Fig. 1], the electric field strength of the incident, reflected and transmitted waves may be written in the following forms [11]:

$$E_0 \exp[in_1 \xi - i\omega t], \tag{7a}$$

$$E_r \exp[-in_1 \xi - i\omega t], \tag{7b}$$

and

$$E_t \exp[in_2(\xi - \phi) - i\omega t], \tag{7c}$$

respectively. From the condition of continuity of the field and its first derivate at the boundaries, we can find:

$$E_0 + E_r = E(0), \tag{8a}$$

$$in_1(E_0 - E_r) = E'(0), \tag{8b}$$

$$E_t = E(\phi), \tag{8c}$$

$$in_2 E_t = E'(\phi), \tag{8d}$$

where $\phi = kl$ and (') stands for the derivative with respect to $\xi$. Introducing the amplitude reflection coefficient $r = E_r/E_0$, we can represent Eqs. (8a) and (8b), as

$$\frac{1-r}{1+r} = \frac{M(0)}{n_1}, \tag{9}$$

where $M(0) = E'(0)/iE(0)$ is the surface admittance of the first boundary of a gas layer. Consequently, reflectivity may be expressed in terms of the surface admittance

$$R = |r|^2 = \left|\frac{n_1 - M(0)}{n_1 + M(0)}\right|^2. \tag{10}$$

According to Eq. (10), in order to study the spectral line profile in reflection, the set of Eqs. (6) needs to be solved with respect to the surface admittance.

### III. FOURIER METHOD: THE CASE OF SPECULAR REFLECTION OF ATOMS COLLIDING WITH THE WALLS

In this section, following the method described in Ref. [12], the exact solution of Eqs. (6a) and (6b) is obtained for the case of specular boundary conditions by means of Fourier series expansion of the field. According to the specular boundary conditions, after a collision with the surface of a dielectric material, an atom conserves its polarization and changes the sign of the velocity component normal to a given surface. Thus, we can set

$$\sigma(\xi=0,\nu)=\sigma(\xi=0,-\nu),\qquad(11a)$$

$$\sigma(\xi=\phi,-\nu)=\sigma(\xi=\phi,\nu),\qquad(11b)$$

for the atoms at the front and rear boundaries. Now, in order to determine the field inside the vapor, let's imagine that there are no boundaries at planes $\xi=0$ and $\xi=\phi$. If we reflect the gas layer with respect to the plane $\xi=0$ and repeat the resulting layer with the thickness of $2\phi$ throughout the space, the field will satisfy the following condition

$$E(j\phi-\xi)=E(j\phi+\xi),\qquad(12)$$

where $j$ is an integer. In this case, the atom reflected from the wall in the layer problem corresponds to an atom freely crossing the interface from the side of the adjacent layer in the infinite space problem.

It is clear from Eqs. (11a), (11b) and (12) that the field and the off-diagonal element of the density matrix can be expanded in the Fourier series as continuous periodic functions with a period of $2\phi$ in the form

$$F(\xi)=\sum_{n=-\infty}^{+\infty}F(n)e^{in\pi\xi/\phi},\qquad(13)$$

where $F(n)=(2\phi)^{-1}\int_{-\infty}^{+\infty}F(\xi)e^{-in\pi\xi/\phi}d\xi$. After we multiply Eqs. (6a) and (6b) by $(2\phi)^{-1}e^{-in\pi\xi/\phi}$ and integrate them from $-\phi$ to $\phi$, the set of equations for the field and the off-diagonal element of density matrix is transformed into

$$(q_n^2-1)E(n)-\phi^{-1}\left[e^{in\pi}E'(\phi)-E'(0)\right]=2im\int_{-\infty}^{+\infty}\sigma(n,\nu)e^{-\nu^2}d\nu,\qquad(14a)$$

$$(\eta+iq_n\nu)\sigma(n,\nu)=E(n),\qquad(14b)$$

where $q_n=n\pi/\phi$. The system of equations (14) can be solved with respect to the Fourier coefficient of the field

$$E(n)=\phi^{-1}\frac{(-1)^n E'(\phi)-E'(0)}{q_n^2-1-2imI(n)},\qquad(15)$$

where $I(n)=\int_{-\infty}^{+\infty}\frac{e^{-\nu^2}}{\eta+iq_n\nu}d\nu$. From Eq. (15) it is seen, that in order to calculate the surface admittance it is essential to find the connection between field derivatives on the layer boundaries. Substitution of Eq. (15) into Eq. (13) yields

$$E(\xi)=\sum_{n=-\infty}^{+\infty}E(n)e^{iq_n\xi}=\sum_{n=-\infty}^{+\infty}\frac{(-1)^n E'(\phi)-E'(0)}{q_n^2-1-2imI(n)}e^{iq_n\xi}.\qquad(16)$$

For the field at the boundaries of the layer $\xi=0$ and $\xi=\phi$, we have

$$E(0)=E'(\phi)S^- - E'(0)S^+,\qquad(17a)$$

$$E(\phi)=E'(\phi)S^+ - E'(0)S^-.\qquad(17b)$$

Here we have introduced two sums

$$S^+ = \phi^{-1} \sum_{n=-\infty}^{+\infty} \frac{1}{q_n^2 - 1 - 2imI(n)}, \qquad (18a)$$

$$S^- = \phi^{-1} \sum_{n=-\infty}^{+\infty} \frac{(-1)^n}{q_n^2 - 1 - 2imI(n)}. \qquad (18b)$$

Finally, solving Eqs. (8c), (8d), and (17) for the ratio $E(0)/E'(0)$, we can find an exact expression for the surface admittance in terms of sums $S^+$ and $S^-$

$$M(0) = \frac{1 - in_2 S^+}{n_2(S^- - S^+)(S^- + S^+) - iS^+}. \qquad (19)$$

It is worth noting that the integrand of $I(n)$ in the denominator of sums [see Eqs. (18a) and (18b)] depends on $n$. This fact necessitates the use of numerical methods for the accurate calculation of reflection spectra. Nevertheless, analytical formulas for the reflection coefficient can be obtained in some limiting cases.

In order to test our theory, let's first assume that there is no vapor in the layer. For this instance, we can set parameter $m$ equal to 0 in Eq. (19). Therefore, we get rid of the integral in Eqs. (18a) and (18b) and sums $S^+$ and $S^-$ converge to explicit analytic functions

$$S^+ = \phi^{-1} \sum_{n=-\infty}^{+\infty} \frac{1}{(n\pi/\phi)^2 - 1} = -\cot\phi, \qquad (20a)$$

$$S^- = \phi^{-1} \sum_{n=-\infty}^{\infty} \frac{(-1)^n}{(n\pi/\phi)^2 - 1} = -\csc\phi. \qquad (20b)$$

Substitution of Eqs. (20a) and (20b) into Eq. (19) leads to the well-known result of the wave interference in the empty Fabry-Perot resonator

$$r = \frac{i(n_1 - n_2) + (n_1 n_2 - 1)\tan\phi}{i(n_1 + n_2) + (n_1 n_2 + 1)\tan\phi}, \qquad (21)$$

where $\phi$ denotes the thickness of the gap between dielectric media $l$ divided by reduced wavelength $\lambdabar = \lambda/2\pi$.

### IV. NUMERICAL CALCULATION

In the following section the reflectivity is calculated numerically using Eqs. (10), (18), and (19) for different thicknesses of the gas layer. The results for eight various thicknesses (from $l = \lambda/2$ to $l = 9\lambda/4$ with a step $\lambda/4$) are presented in Fig. 2, where the reflectivity is plotted as a function of dimensionless detuning $\Omega$.

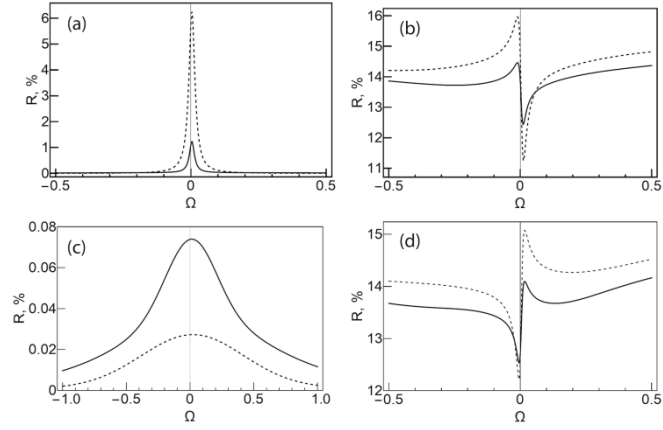

FIG. 2. Reflectivity of a thin vapor layer against dimensionless detuning for eight different layer thicknesses. Dashed curves: a – $l=\lambda/2$, b – $l=3\lambda/4$, c – $l=\lambda$, d – $l=5\lambda/4$; Solid curves: a – $l=3\lambda/2$, b – $l=7\lambda/4$, c – $l=2\lambda$, d – $l=9\lambda/4$. The parameters $\Gamma$, $n_1$, $n_2$, and $m$ were taken equal to 0.01, 1.5, 1.5, and 0.001, respectively. Equations (10), (18) and (19) were used to calculate the reflectivity.

From the comparison of dashed and solid curves in Fig. 2, it can be seen that with an increase in the layer thickness by the wavelength of the incident light, the spectral line contour of the reflection coefficient almost periodically repeats itself. This $\lambda$-periodic dependence is derived from spatial oscillations of the transient component of the polarization. The conditions $l=(2n-1)\lambda/2$ and $l=(n-1/2)\lambda/2$, where $n=1,2,3,...$, correspond to the presence of the sub-Doppler structure with approximately even (with Lorentzian line shape) [see Fig. 2(a)] and odd [see Fig. 2(b) and (d)] spectral contours (with respect to the resonant transition of the gas atoms), respectively.

Apart from that, the presence of the Fabry-Perot resonances determines the $\lambda/2$-periodic dependence of the non-selective contribution to the total reflection on the layer thickness. It is also worth noting that in the limit of large thicknesses of the gas layer, when $l$ significantly exceeds the wavelength of the incident light, the second boundary ceases to contribute to total reflection due to the attenuation of the transmitted wave. Consequently, numerical calculation gives the well-known spectral profile of the selective reflection of light from a thick gas layer with the logarithmic singularity (this result is presented in Fig. 3).

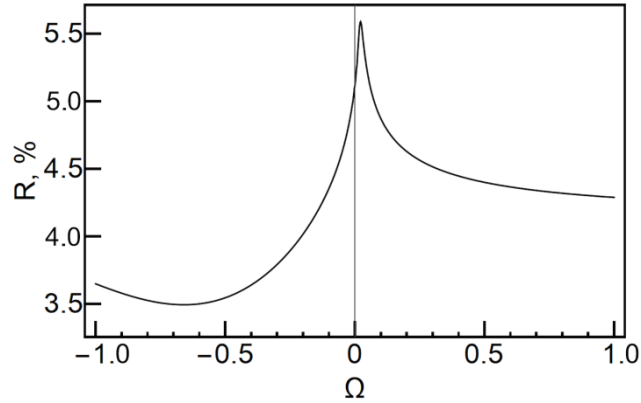

FIG. 3. Reflectivity of a thick gas layer in the resonance region versus dimensionless detuning. The parameters $\Gamma$, $n_1$, $n_2$, $m$, and $l$ were taken equal to 0.01, 1.5, 1.5, 0.005, and $500\lambda$, respectively. Equations (10), (18) and (19) were used to calculate the reflectivity.

Since in the first order in vapor density, the position of the even Doppler-free peak in the selective reflection spectra coincides with the atomic transition frequency $\omega_0$ [8], to study the "blueshift" it is necessary to consider the high-order terms with respect to $m$. Equations (10), (18), and (19) allow us to accurately calculate the values of the "blueshift" in a wide range of atomic number densities. With an increase in the concentration of atomic vapors, the spectral line contour of the reflection coefficient broadens, the amplitude of the maximum increases, and its position shifts to the short-wavelength part of the spectrum. Because of the inequality $m < \Gamma$ that holds due to the self-broadening effect [14] in the domain of the strong spatial dispersion $\Gamma < 1$, the shift of the resonance frequency $\Delta\Omega$ is well described by a linear dependence on $m$. In the case of a thick layer a universal proportionality constant may be found in the framework of the perturbation theory [8]. Contrary to that, in the case of a thin gas layer the meaning of the shift is to be elaborated because the line shape depends on the layer thickness. As an example, we quote here the relation $\Delta\Omega = 3.54m$ obtained numerically in the case of an even spectral line corresponding to $l = \lambda/2$ and $\Gamma = 0.01$.

## V. DISCUSSION AND CONCLUSION

The special role of the cavities of half wavelength thickness in the Doppler width cancellation was first pointed out in the microwave region. The main idea of [15] was that in a pillbox shaped cell the slowly moving atoms contribute most to the absorption, as well as to the fluorescence, simply because they spend more time in their free flights between the walls. With the highly reflective walls the field inside the cell is very close to a standing wave. Then, in the case of a half wavelength thick cell all atoms experience the field oscillations of the same phase. Nevertheless, after the work on selective reflection from a gaseous half-space [1] it becomes clear that the main reason for the Doppler width cancellation is the transient polarization of the atoms that was not accounted for in [15]. Interference of two transient contributions that start their oscillations from the opposite walls leads to the periodic dependence of the line shapes on the cell thickness and further enhancement of the effect [3], while the details of the scattering at the walls are of minor importance. Because of that, the results of the rigorous solution of the problem formulated for the specular reflection of the atoms have the general significance and may be used as a clue also in the case of diffuse and quenching collisions for which the rigorous solutions are still absent. Below we specify some peculiarities connected with the specular reflection of atoms colliding with the walls.

We can see from the above that the amplitude of selective reflection of light from a thin layer of resonant gas has unexpectedly large values in the case of specular reflection of atoms colliding with the walls. Since at room temperatures the interaction of atoms with a dielectric surface is mostly of diffuse type, it is important to underline the main differences between models of diffuse and specular reflection. As shown in Ref. [13], at large layer thicknesses, the spectral line profile of the selective reflection signal has a logarithmic form, and its amplitude for the case of specular reflection is approximately two times larger than the corresponding amplitude in the case of diffuse boundary conditions. However, if the thickness of the vapor layer is comparable to the wavelength $\lambda$, the differences in the amplitude and line shape of the resonance for two models are much more considerable.

In the case of specular reflection, collisions with the walls allow a large part of atoms to oscillate at their own frequency for a long time. The interference of contributions from individual atoms leads to the formation of a narrow and intense peak with a Lorentzian spectral profile near the resonance. With an increase in the layer thickness, atoms have enough time to adapt to the external field before they collide with the wall. As a result, the amplitude of the Lorentz spectral contour decreases significantly and the spectral contour of selective reflection has a mostly logarithmic singularity [see Fig. 3]. It is also worth noting that the condition $l = n\lambda$ corresponds to the absence of sub-Doppler structure. In this case, the Doppler-wide spectral profile [see Fig. 2(c)] arises due to the absorption of the light in a gaseous medium.

Finally, the shift of the resonant frequency is reasonably well described by a linear dependence on the number density of gas atoms. This result is consistent with the magnitude of the "blueshift" obtained analytically in Ref. [13] for the case of a thick gas layer and specular wall collisions. Although, it was shown numerically in [9] and analytically in [8] that the Lorentz-Lorenz local-field correction partially reduces the value of the "blueshift", this shift of purely electrodynamics nature is to be accounted for in any attempts of measuring van der Waals atom-surface interaction in selective reflection from and transmission through thin vapor layers.